# Spin-chirality-dependent modulation of topological gap, Chern number, and valley-polarization in monolayer kagome lattice $Cr_3Se_4$

Wenzhe Zhou[1], Guibo Zheng[1], Yating Li[1], Aolin Li[2], and Fangping Ouyang[1,2,3,*]

1*School of Physics, Hunan Key Laboratory for Super-Microstructure and Ultrafast Process, and Hunan Key Laboratory of Nanophotonics and Devices, Central South University, Changsha 410083, People's Republic of China*

2*School of Physics and Technology, State Key Laboratory of Chemistry and Utilization of Carbon-Based Energy Resources, Xinjiang University, Urumqi 830046, People's Republic of China*

3*State Key Laboratory of Powder Metallurgy, and Powder Metallurgy Research Institute, Central South University, Changsha 410083, People's Republic of China*

## Abstract

Kagome materials exhibit unique electronic properties, such as the quantum anomalous Hall effect. The control of Chern numbers is critical for quantum device manipulation, but existing research has mainly focused on collinear magnetization while neglecting chiral spin textures. Through first-principles calculations and tight-binding modeling of monolayer $Cr_3Se_4$, this study reveals spin-chirality-dependent control of topological gaps, Chern numbers, and valley polarization in kagome materials. The results demonstrate that the azimuthal angle has no observable effect. For collinear magnetization ($\kappa$ = 0) or spin-chirality $\kappa$ = -1, the topological bandgap decreases as the spin orientation approaches the in-plane direction. Conversely, increasing the polar angle enhances the bandgap for $\kappa$ = 1. In the breathing kagome lattice, the degeneracy between K and K' valleys is lifted. As the gap undergoes sequential closure and reopening in the two valleys, the structural asymmetry and spin-chirality allow for controlled tuning of the topological gap, Chern number, and valley polarization. Moreover, the emergence of a topological Hall effect is is also demonstrated. These findings provide strategies for controlling topological states and advancing applications in quantum devices and valleytronic systems.

*Corresponding author. E-mail address: ouyangfp06@tsinghua.org.cn

The quantum anomalous Hall effect (QAHE) has emerged as a cornerstone of topological matter research, bridging fundamental concepts in band topology with fault-tolerant quantum device architectures. Numerous two-dimensional magnetic materials have been theoretically identified as potential platforms for realizing the QAHE. Cr doped $(Bi, Sb)_2Te_3$ was first observed in experiment for QAHE[1]. The QAHE is defined by a quantized Hall conductance $\sigma_{xy} = C e^2/h$ at zero magnetic field[2], where $C$ is the integer Chern number. The corresponding insulator is called a Chern insulator. The quantized Chern number, as a topological invariant, governs the transition between distinct quantum phases, making its controlled manipulation a central challenge in condensed matter physics[3,4].

The Chern number can be calculated by integrating the Berry curvature in reciprocal space, where the Berry curvature is typically localized near band valleys. Valley has garnered significant attention as a novel degree of freedom[5-8]. In hexagonal systems with broken inversion symmetry, the K and -K valleys exhibit opposite Berry curvatures, giving rise to valley-contrasting phenomena. In ferromagnetic materials with broken time-reversal symmetry, the energy degeneracy of inequivalent valleys can be lifted, a class of systems now termed ferrovalley materials[9,10]. Many materials with symmetry breaking are predicted to have ferrovalley properties, such as $MX_2$ (M = V, Nb, Gd, Ru, Y; X = S, Se, Br, I)[11-15], $VSi_2N_4$[16], $CuCrP_2S_6$[17], $Nb_3I_8$[18], and their Janus compounds[19-22]. Moreover, spin-valley coupling in antiferromagnetic materials induce intrinsic valley polarization despite spin degeneracy[23,24]. Valley polarization is intrinsically linked to Berry curvature, where the opening and closing of valley gaps can be modulated by external fields, giving rise to valley-dependent QAHE[25-29]. Magnetic order serves as a crucial means to modulate valley-dependent properties[30,31]. However, existing studies predominantly focus on valley polarization and QAHE in collinear magnetic systems, with limited consideration of spin-chirality.

Spin-chirality provides an additional degree of freedom for quantum control. Non-collinear spin textures are frequently observed in kagome lattices, including $Mn_3X$ (X = Rh, Ir, Pt)[32,33] and $Mn_3Y$ (Y = Ge, Sn, Ga)[34,35]. The kagome lattices, composed of corner-sharing triangles and hexagons, exhibits distinctive electronic properties including superconductivity, charge density waves, and anomalous Hall effects. Despite significant experimental challenges in fabricating two-dimensional kagome systems, their unique structural and electronic properties continue to attract considerable interest.[36]. Materials such as $Co_3X_3Y_2$ (X = C, Si, Ge, Sn, Pb; Y = O, S, Se, Te, Po)[37,38], $Cr_3Te_4$[39], $Cr_3O_4Cl$[40], their Janus compounds[41], and other two-dimensional kagome materials have been predicted to exhibit quantum anomalous Hall effect[42]. Valley polarization and ferroelectric transitions in breathing kagome materials with symmetry breaking have also garnered widespread attention[43-45]. The valley-dependent topological transitions in breathing kagome lattices were also predicted[46]. Zhou et al. employed a kagome material as a magnetic substrate to investigate the modulation of the Chern number by spin-chirality , which involves the effect of *d*-electron spin-chirality on s-electrons[47]. Since spin and orbital degrees of freedom are coupled, the spin-chirality of *d*-electrons may impose distinct modulation effects on QAHE contributed by *d*-orbitals, which warrants further investigation.

In this work, we theoretically demonstrate spin-chirality-dependent modulation of topological bandgaps and Chern numbers based on a tight-binding model of kagome lattices. A tight-binding model was constructed using the orbital contributions to the intrinsic quantum anomalous Hall effect in monolayer $Cr_3Se_4$ as a representative example. The combination of symmetry breaking and spin-chirality enables broader tunability of Chern numbers. Our findings significantly expand the methodology for Chern number modulation in quantum anomalous Hall systems, providing crucial guidance for experimental realization.

The first-principles calculations on the electronic structure of monolayer $Cr_3Se_4$ have been performed by using the plane-wave basis first-principles package DS-PAW software, which is integrated into the visualization software Device Studio[48]. The projector augmented wave (PAW) method and the Perdew-Burke-Ernzerhof (PBE) exchange-correlation functional with the generalized gradient approximation (GGA) are used[49,50]. The cut-off energy of the plane-wave basis was set to 600 eV. The criteria for force and energy are 0.01 eV/Å and $10^{-6}$ eV, respectively, and a $15 \times 15 \times 1$ Γ-centered k-point grid was used for k-point sampling. A vacuum slab of at least 20 Å is introduced along the z-direction. To treat localized 3*d* electrons of Cr atoms, the GGA+U method is employed with $U_{eff}$ = 3 eV[51].

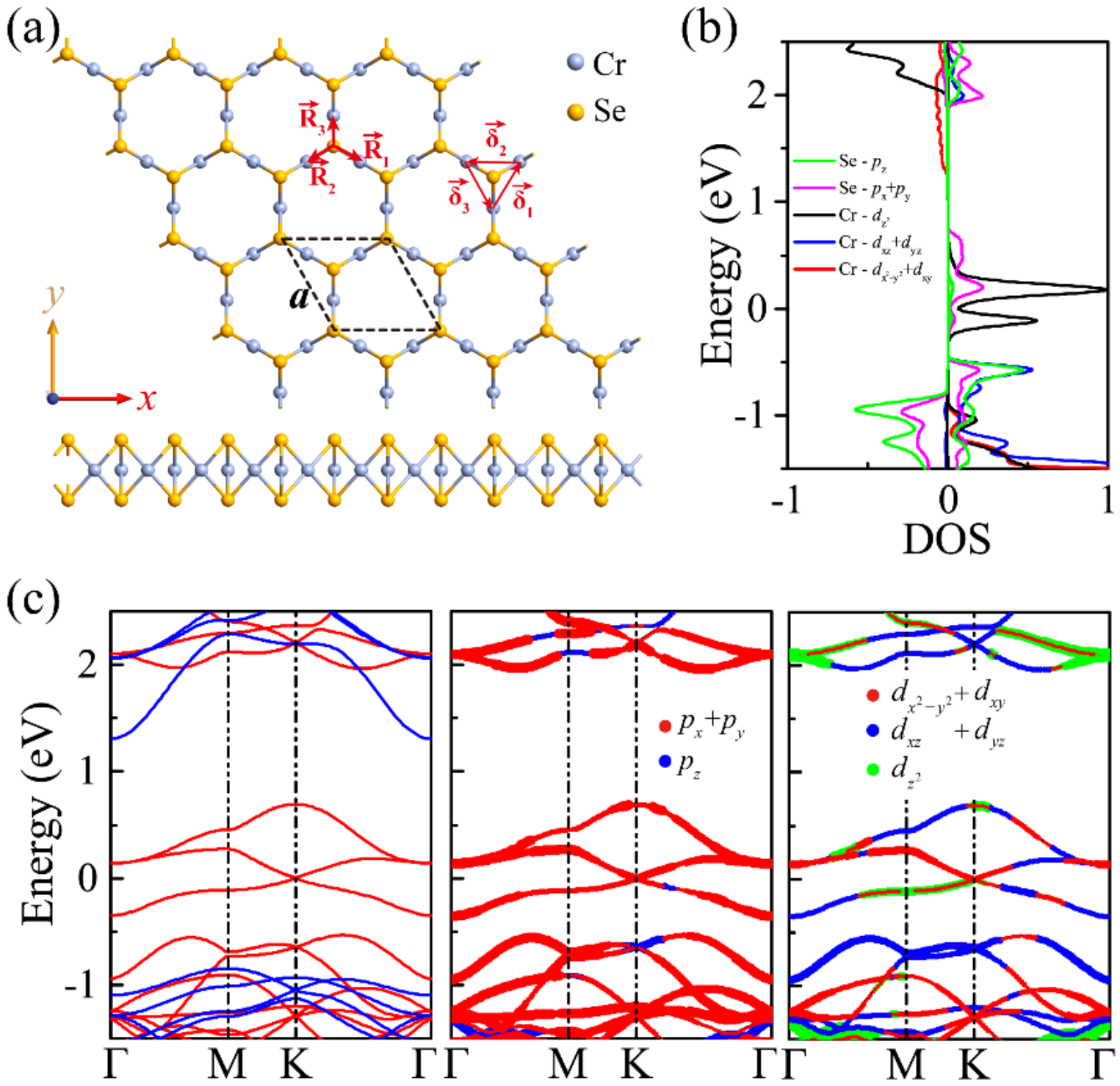

**Fig. 1** (a) Top and side views of monolayer kagome $Cr_3Se_4$. (b) The spin-resolved projected density of states for monolayer $Cr_3Se_4$. (c) The band structures of monolayer $Cr_3Se_4$ without spin-orbit coupling. The left panel shows spin-resolved band structures, where red and blue colors represent spin-up and spin-down states, respectively. The middle panel is the band structure projected onto the *p* orbitals of Se atoms. The right panel is the band structure projected onto the *d* orbitals of Cr atoms. The density of states and band structures are obtained by the first-principles calculations utilizing the hardware and software resources of HZWTECH[48].

The structure of monolayer $Cr_3Se_4$ is shown in **Fig. 1(a)**. Cr atoms form a kagome lattice, with its atomic layer sandwiched between two layers of Se atoms. Each Se atom interacts with three Cr atoms, and each Cr atom is nearest to four Se atoms. The optimized lattice constant of monolayer $Cr_3Se_4$ is 6.25 Å. The stability has been confirmed in previous research[52]. First-principles calculations identify monolayer $Cr_3Se_4$ as a spin-polarized Dirac semi-metal, with the Dirac point at the Fermi level of the K point, as shown in **Fig. 1(c)**. The orbital-projected band structures and the density of states indicate that the Dirac point is contributed by the $d_{x^2-y^2}$, $d_{xy}$, $d_{z^2}$ orbitals of Cr atoms and the $p_x$, $p_y$ orbitals of Se atoms. Therefore, the tight-binding model of the kagome lattice can be constructed based on these five orbital basis vectors.

The tight-binding Hamiltonian of monolayer $Cr_3Se_4$ is

$$H = \sum_{i\alpha} \varepsilon_{i\alpha}^{(p)} p_{i\alpha}^{\dagger} p_{i\alpha} + \sum_{j\alpha} \varepsilon_{j\alpha}^{(d)} d_{j\alpha}^{\dagger} d_{j\alpha} + \sum_{\langle\alpha,\beta\rangle,\vec{\delta}_i} t_{\alpha\beta}^{(\vec{\delta}_i)} d_{\alpha}^{\dagger} d_{\beta}^{(\vec{\delta}_i)} + \sum_{\langle\alpha,\beta\rangle,\vec{R}_i} t_{\alpha\beta}^{(\vec{R}_i)} p_{\alpha}^{\dagger} d_{\beta}^{(\vec{R}_i)} + \sum_{j\alpha} \vec{m}_{j\alpha} \bullet \vec{\sigma} + \lambda \vec{L} \bullet \vec{\sigma} + h.c. , \quad (1)$$

where $\alpha$ and $\beta$ denote the two $p$ orbitals and three $d$ orbitals. The first two terms are the on-site terms of $p$ orbitals and $d$ orbitals, respectively. The third term is the nearest-neighbor hopping between $d$ orbitals, and the fourth term is the nearest-neighbor hopping between $d$ orbitals and $p$ orbitals. The fifth and sixth terms represent the effective spin splitting introduced by the magnetic moment and the spin-orbit coupling effect. $\vec{\sigma}$ is the Pauli matrix for spin, and $\vec{L}$ is the operator of the orbital angular momentum. $\vec{m}_{j\alpha}$ represents the magnetic direction for the $\alpha$ orbital at the $j$ site. $\lambda$ is the strength of spin-orbit coupling. Based on the first-principles calculations, we selected the $d_{x^2-y^2}$, $d_{xy}$, $d_{z^2}$, $p_x$ and $p_y$ orbitals as the basis vectors. More details about the Hamiltonian matrix and the calculations of Berry curvature and Chern number can be seen in the Supplementary Materials (SM).

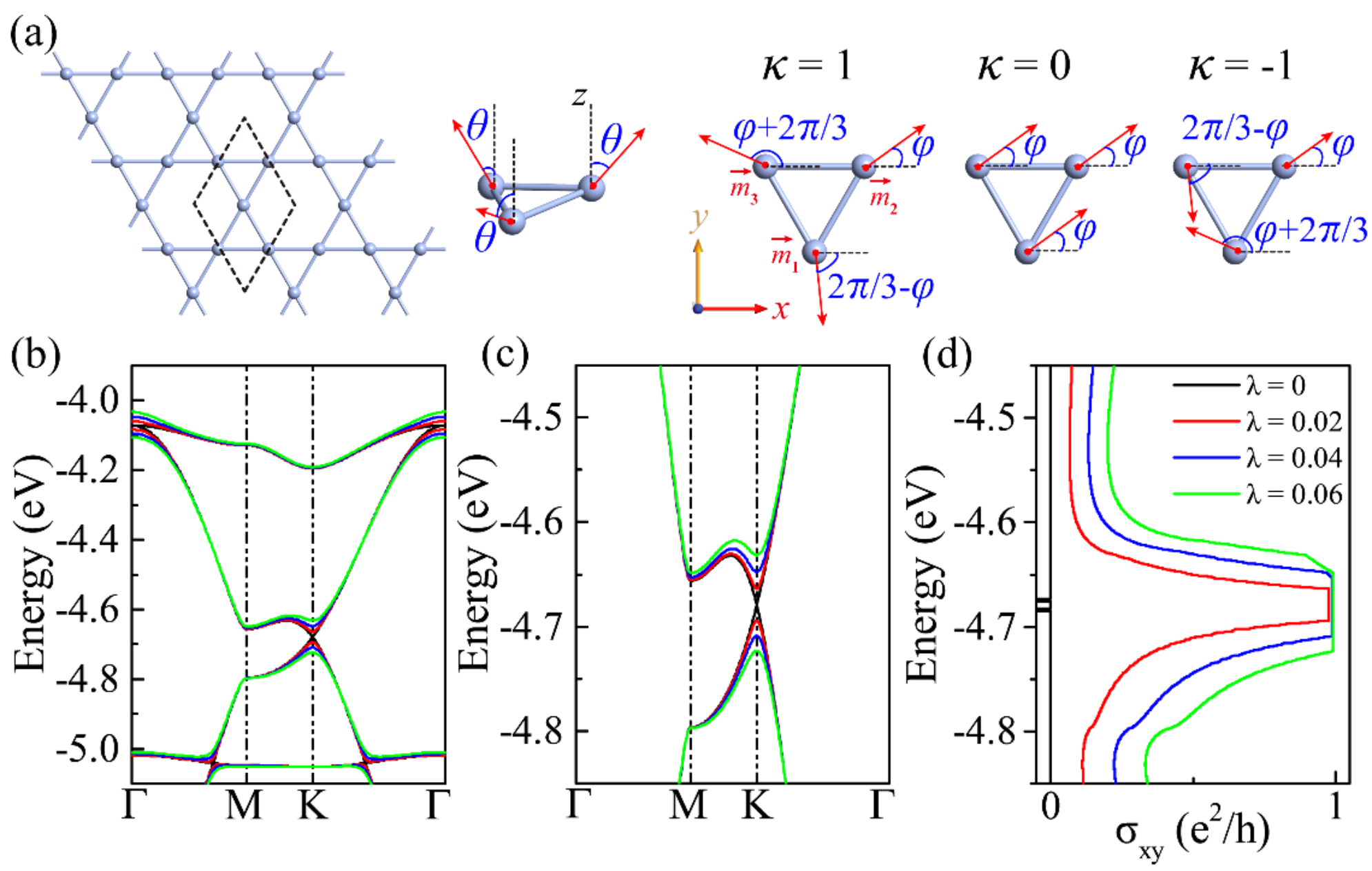

**Fig. 2** (a) The illustration of the kagome lattice and the magnetic directions of the magnetic atoms in the unit-cell, labeled with the angles $\theta$, $\varphi$, and the chirality $\kappa$. (b) The band structures of the tight-binding kagome model with different strengths of spin-orbit coupling. (c) The magnified image of the band structures shows the topological band. (d) The anomalous Hall conductance at different spin-orbit coupling strengths. The chirality $\kappa = 0$ and the angle $\theta = 0$.

The spin-chirality is illustrated in **Fig. 2(a)**, where the spin configuration can be described by the chirality $\kappa$, polar angle $\theta$, and azimuthal angle $\varphi$. The polar angles of the three Cr atoms in a unit of the kagome lattice are the same, but the azimuthal angles differ by 120°. The positive ($\kappa = 1$) and negative ($\kappa = -1$) chiralities are related to the anticlockwise and clockwise rotations of the azimuthal angle. For $\kappa = 0$, the magnetic order is the collinear ferromagnetic order.

The band structure of the constructed tight-binding kagome model is shown in **Fig. 2(b, c)**. When the strength of spin-orbit coupling is zero, there is a Dirac state at the K point. With the spin-orbit coupling effect, the Dirac point opens an energy gap. The calculated anomalous Hall conductivity reveals a quantized conductance plateau, demonstrating the topological nature with a Chern number $C = 1$. The topological gap at the K point increases with the strength of spin-orbit coupling. The modulation of topological gap and Chern number of the kagome model with collinear ferromagnetic order ($\kappa = 0$) is demonstrated in the SM. The modulation behavior shows excellent agreement with the predictions in the literature[39], demonstrating the validity of the constructed model.

The influence of spin-chirality on the topological gap and Chern number is investigated. The azimuthal angle $\varphi$ is irrelevant to the electronic structure. For the chirality $\kappa = 1$, as the polar angle $\theta$ is tuned from 0° to 90°, the topological gap at the K point gradually increases from 0.061 eV to 0.348 eV. The Berry curvature at the K and K' points gradually diminishes and becomes non-local as the gap increases, eventually vanishing when $\theta = 90°$, as shown in **Fig. 3(c)**. The evolution of Berry curvature and Chern number from $\theta = 90°$ to 180° mirrors that from 0° to 90° with opposite sign, with $\theta = 90°$ serving as the critical point for topological phase transition.

For chirality $\kappa$ = -1, the topological gap at the K point exhibits a non-monotonic evolution—first decreasing and then increasing—as the polar angle $\theta$ is continuously tuned from 0° to 90°. At the critical angle $\theta$ = 32.448°, the system undergoes a band gap closing. Concurrently with the gap reduction, the Berry curvature at the valley demonstrates progressively enhanced localization. Notably, a discontinuous topological phase transition occurs at the critical angle, manifested by an abrupt reversal of the Berry curvature from a positive maximum to a negative maximum. As $\theta$ increases beyond the critical angle toward 90°, the absolute value of the Berry curvature rapidly decays and eventually vanishes. At the critical angle, the Chern number exhibits a quantized transition from +1 to -1, signaling a topological phase change. Remarkably, at $\theta$ = 90°, the system undergoes another topological transition where the Chern number reverts from -1 back to +1. $\theta$ = 147.552° is also a critical angle, as clearly demonstrated in **Fig. 3(g-h)**, where similar topological characteristics emerge.

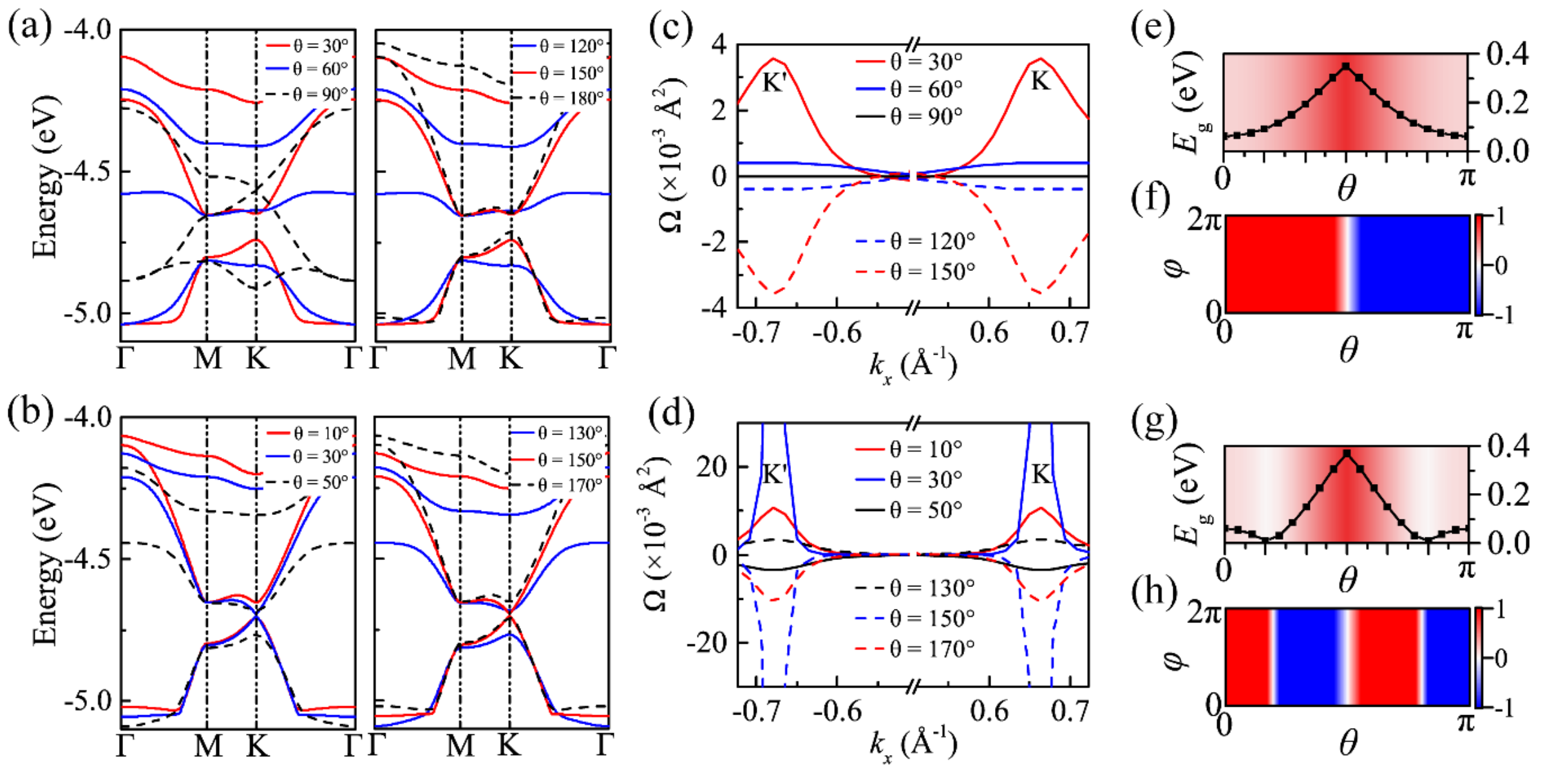

**Fig. 3** The band structures of the kagome model with different angles of $\theta$, for chirality $\kappa$ = 1 (a) and $\kappa$ = -1 (b). The Berry curvatures along the $k_x$ direction ($k_y$ = 0), for chirality $\kappa$ = 1 (c) and $\kappa$ = -1 (d). The variation of the topological gap at the K point with the angle $\theta$, for chirality $\kappa$ = 1 (e) and $\kappa$ = -1 (g). The variation of the Chern number with the angle $\theta$, for chirality $\kappa$ = 1 (f) and $\kappa$ = -1 (h). The spin-orbit coupling strength $\lambda$ is 0.04 eV.

The topological gap and Berry curvature at K and K' points evolve synchronously. To introduce additional controllable degrees of freedom, the magnetic-order-dependent electronic structure in the breathing kagome lattice is investigated. The two types of corner-sharing triangles exhibit distinct sizes, leading to different hopping integrals as illustrated in **Fig. 4(a)**. For simplicity, we scale the hopping integrals within the triangles by a factor $t'$ relative to their original values. For $t' = 1$, it is the symmetric kagome lattice. For $t' \neq 1$, the central inversion symmetry is broken. With the breaking of inversion symmetry, K and K' are not degenerate, which is consistent with previous reports[46], as shown in **Fig. 4(b)**. As the parameter $t'$ changes from 0.8 to 1.2, the energy gaps at K and K' points both decrease linearly first to zero and then increase. The greater the deviation of $t'$ from 1 (indicating a larger structural difference between the two triangles in the breathing kagome lattice), the more pronounced the gap difference between the K and K' valleys becomes, as illustrated in **Fig. 4(d)**. When $t'$ is approximately 0.95 and 1.05, the energy gaps at K' and K points are zero. At these critical points, topological transitions occur as the Berry curvatures at the K' and K points successively change sign. When $t'$ is between 0.95 and 1.05, the Chern number of the system is 1. When $t'$ is less than 0.95 or greater than 1.05, the Chern number of the system is 0.

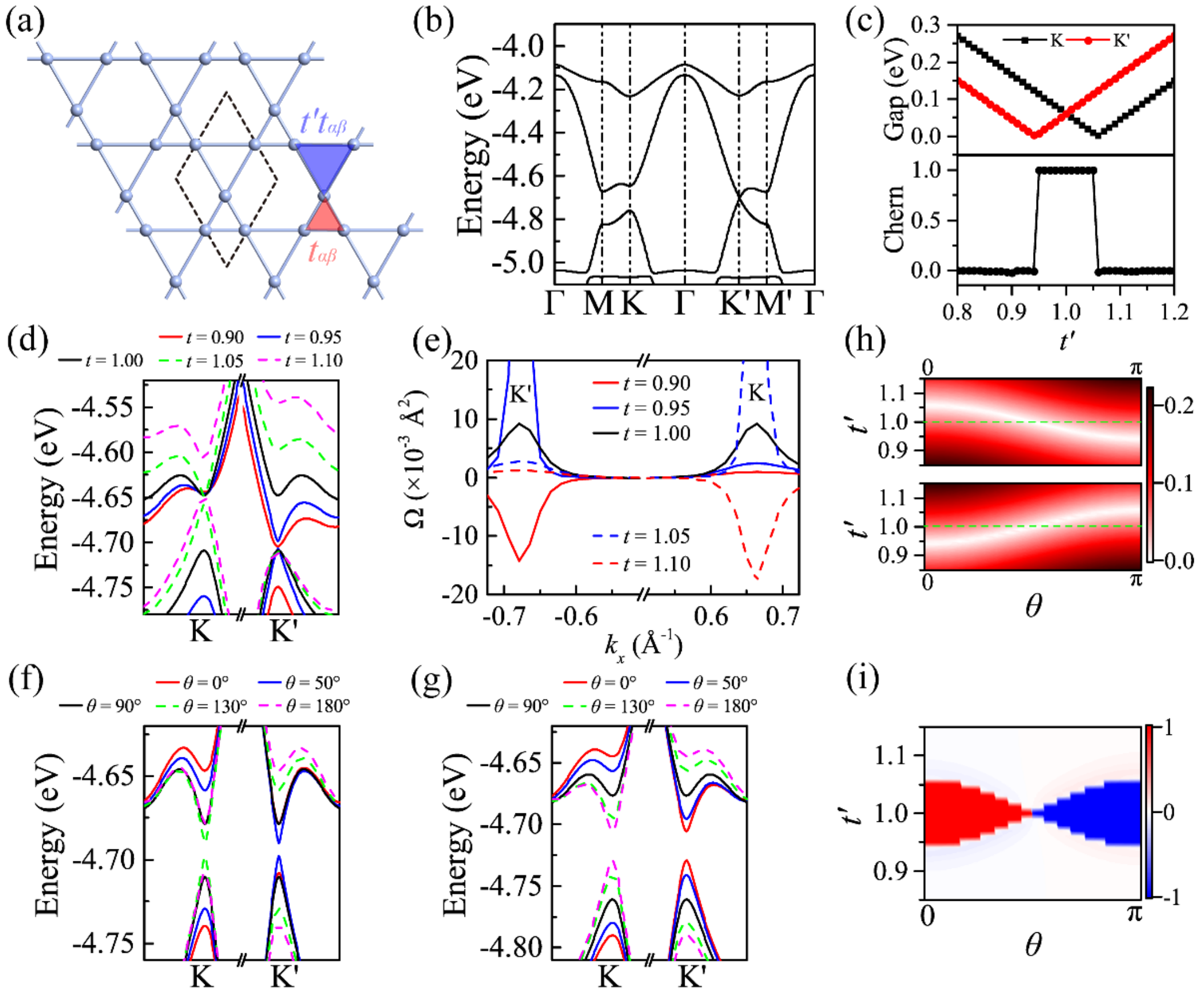

**Fig. 4** (a) The illustration of breathing kagome lattice. (b) The band structures of the breathing kagome lattice with the parameter $t' = 0.95$, where $\kappa = 0$, $\theta = 0$, $\varphi = 0$. (c) The topological gaps at K and K' points (upper) and the Chern number of the model (lower) change with the parameter $t'$, where $\kappa = 0$, $\theta = 0$, $\varphi = 0$. The band structures (d) and Berry curvatures along the $k_x$ direction (e) of the breathing kagome model with different parameter $t'$, where $\kappa = 0$, $\theta = 0$, $\varphi = 0$. The band structures of the breathing kagome model with different $\theta$ for $t' = 0.97$ (f) and $t' = 0.92$ (g), where $\kappa = 0$, $\varphi = 0$. The topological gaps (h) at K (upper) and K' points (lower) and the Chern number (i) dependent on the angle $\theta$ and parameter $t'$ with chirality $\kappa = 0$ and $\varphi = 0$. The spin-orbit coupling strength $\lambda$ is 0.04 eV.

For the chiral $\kappa = 0$, as the polar angle varies from 0° to 90°, the band gap further decreases. The band structures of the breathing kagome model with different θ for $\kappa = 0$ are shown in **Fig. 4(f, g)**. When $t' < 0.95$, after the K' valley has undergone gap closure and reopening with a Chern number of 0, the continued reduction of the band gap leads to a decrease in the K valley gap while increasing the K' valley gap, with the Chern number remaining at 0. When $0.95 < t' < 1$, the band gap of the K' valley is smaller than that of the K valley. As the band gap decreases, the K' valley first undergoes gap closure followed by reopening, resulting in a Chern number of 0. Subsequently, the K valley experiences gap closure and reopening, leading to a Chern number of -1. **Fig. 4(h, i)** present the topological gaps and Chern numbers modulated by $t'$ and $\theta$ for the collinear ferromagnet.

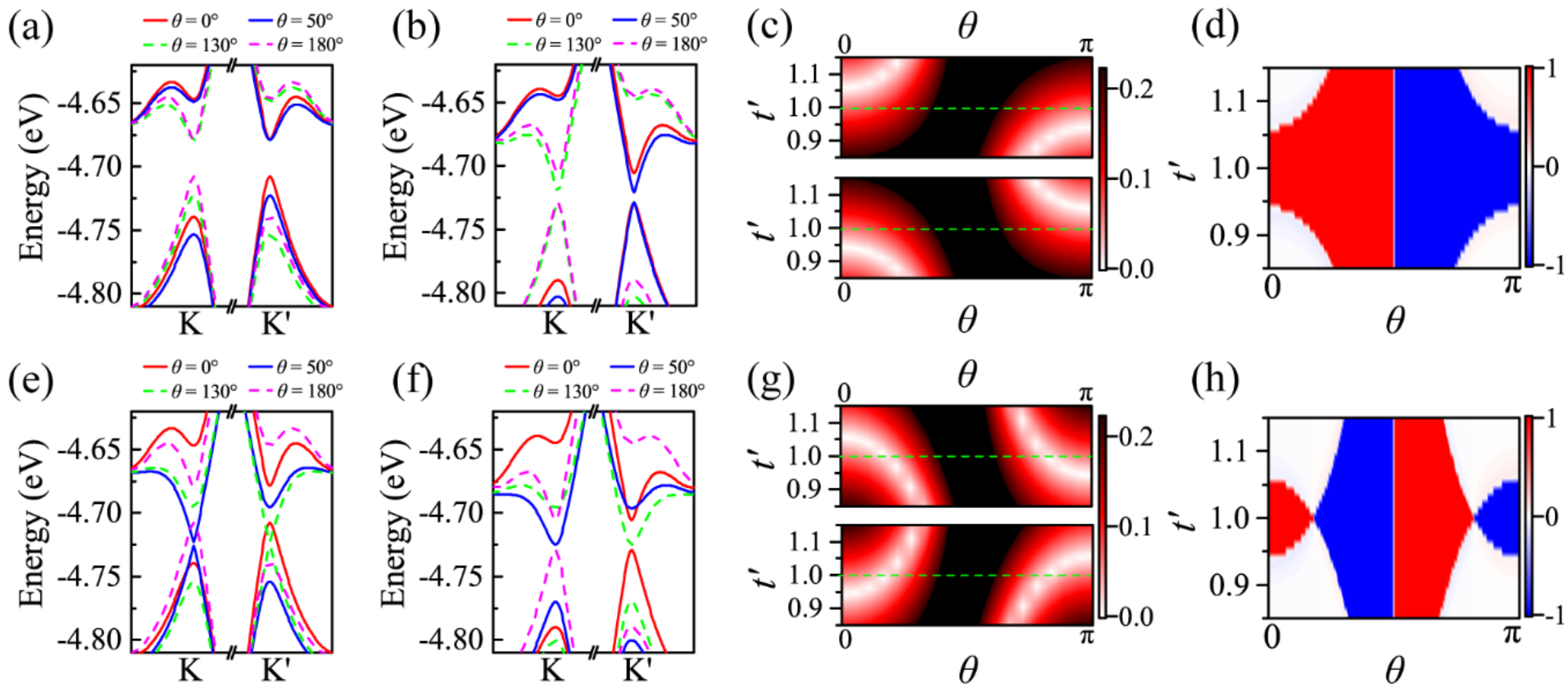

**Fig. 5** The band structures of the breathing kagome lattice with different $\theta$ for $t' = 0.97$ (a) and $t' = 0.92$ (b), where $\kappa = 1$, $\varphi = 0$. The topological gaps (c) at K (upper) and K' points (lower), and the Chern number (d) of the model change with the parameter $t'$, where $\kappa = 1$, $\varphi = 0$. The band structures of the breathing kagome lattice with different $\theta$ for $t' = 0.97$ (e) and $t' = 0.92$ (f), where $\kappa = -1$, $\varphi = 0$. The topological gaps (g) at K (upper) and K' points (lower), and the Chern number (h) of the model change with the parameter $t'$, where $\kappa = -1$, $\varphi = 0$.The spin-orbit coupling strength $\lambda$ is 0.04 eV.

As $\theta$ sweeps from 0° to 180°, the topological gap decreases for $\kappa = 0$ or -1 but increases for $\kappa = 1$. When a valley experiences the closing and reopening of the gap, its Berry curvature reverses, and the Chern number becomes zero. When both valleys undergo gap closure and reopening, the Chern number inverts its sign.

**Fig. 5** illustrates the spin-chirality-dependent modulation of gaps and Chern number. For the chirality $\kappa = 1$, when the parameter $t'$ ranges from 0.95 to 1.05, further increase of the gap neither leads to gap closure nor reopening, thus preventing topological transitions except at $\theta = 90°$. However, for $t' > 1.05$ or $t' < 0.95$, increasing the gap induces repeated closing and reopening of the pre-existing gaps, driving a topological phase transition characterized by the Chern number switching from zero to nonzero values. For the opposite chirality ($\kappa = -1$), when the parameter $t'$ ranges from 0.95 to 1.05, the gap reduction induces successive closure and reopening of both valleys. This drives a topological phase transition in the breathing kagome lattice from a quantum anomalous Hall state to a normal state and back to the quantum anomalous Hall state. For $t' > 1.05$ or $t' < 0.95$, the structural asymmetry induces successive closing and reopening of the valleys: first one valley undergoes gap closure-reopening, followed by the other valley as $\theta$ increases, ultimately resulting in a Chern number of -1.

Notably, in the case of non-collinear magnetic order, spin-orbit coupling only marginally enhances the band gap opening. It is predictable that noncollinear spins can generate a topological electronic structure even without spin-orbit coupling, which is also known as the topological Hall effect[47]. As shown in the SM, spin-chirality affects the topological gap and Chern number even without spin-orbit coupling, indicating the presence of a topological Hall effect.

In summary, monolayer kagome material $Cr_3Se_4$ exhibits quantum anomalous Hall effects. Through the construction of both symmetric and asymmetric kagome lattice models, we systematically investigated the influence of spin-chirality on topological bandgaps, Chern numbers, and valley polarization properties. The results demonstrate that the azimuthal angle $\varphi$ plays no role in the modulation. For collinear magnetic order or $\kappa$ = -1 chirality, the band gap decreases as the polar angle $\theta$ increases from 0° to 90°, whereas for $\kappa$ = 1 chirality, the band gap increases. The asymmetry increases the gap of one valley while reducing the gap of another valley, with the polar angle $\theta$ exerting identical modulation on both gaps. The closing and reopening of K and K' valley gaps induce Berry curvature reversal, thereby regulating the Chern number. The combination of breathing kagome lattice parameter $t'$ and spin-chirality enables broader-range modulation of band gaps, Chern numbers, and valley polarization. In addition, the topological Hall effect arisen from noncollinear spins without spin-orbit coupling is also presented. Our work presents a new platform for tuning the QAHE and Chern number based on chiral spin textures, rather than collinear magnetization.

**Supplementary Material**

See the online supplementary material for details of the tight-binding calculations of the magnetic Kagome lattice, the modulation of topological gap and Chern number of the Kagome model with collinear ferromagnetic order, and the effects of spin-chirality on topological gap and Chern number without spin-orbit coupling.

**Acknowledgment**

*This work is financially supported by the National Natural Science Foundation of China (Grant No. 52073308, No. 12004439, No. 12164046, and No. 12304097), the Key Project of the Natural Science Program of Xinjiang Uygur Autonomous Region (Grant No. 2023D01D03), the Tianchi-Talent Project for Young Doctors of Xinjiang Uygur Autonomous Region (No. 51052300570), the Outstanding Doctoral Student Innovation Project of Xinjiang University (No. XJU2023BS028), Hunan Provincial Natural Science Foundation of China (Grant No. 2023JJ40703), and the State Key Laboratory of Powder Metallurgy at Central South University. This work was carried out in part using computing resources at the High-Performance Computing Center of Central South University. We also gratefully acknowledge HZWTECH for providing computational facilities and technical support.*